# Teacher-student relationship and teaching styles in primary education. A model of analysis.

María-Eugenia Cardenal[a], Octavio-David Díaz-Santana[b] and Sara-María González-Betancor[c]

[a] Universidad de Las Palmas de Gran Canaria, Spain, ORCID: 0000-0001-7797-2711; [b] Universidad de Las Palmas de Gran Canaria, Spain, ORCID: 0000-0002-8197-0214; [c] Universidad de Las Palmas de Gran Canaria, Spain, ORCID: 0000-0002-2209-1922; corresponding author: sara.gonzalez@ulpgc.es

**Abstract**

Purpose:

The teacher role in the classroom can explain important aspects of the student's school experience. The teacher-student relationship, a central dimension of social capital, influences students' engagement, and the teaching style plays an important role in student outcomes. But there is scarce literature that links teaching styles to teacher-student relationship. This article aims to: 1) analyze whether there is a relationship between teaching styles and the type of relationship perceived by students; 2) test whether this relationship is equally strong for any teaching style; and 3) determine the extent to which students' perceptions vary according to their profile.

Design/methodology/approach:

A structural equation model with four latent variables is estimated: two for the teacher-student relationship (emotional vs. educational) and two for the teaching styles (directive vs. participative), with information for 21126 sixth-grade primary-students in 2019 in Spain.

Findings:

- Teacher-student relationships and teaching styles are interconnected.
- The participative style implies a better relationship.
- The perceptions of the teacher are heterogeneous, depending on gender (girls perceive clearer than boys) and with the educational background (children from lower educational background perceive both types of teaching styles more clearly).

Originality/value:

The analysis is based on the point of view of the addressee of the teacher's work, i.e. the student. It provides a model that can be replicated in any other education system.

The latent variables, based on a periodically administered questionnaire, could be estimated with data from diagnostic assessments in other countries, which in turn would allow the formulation of context-specific educational policy proposals that take into account student feedback.

**Abstract**

Purpose:

The teacher role in the classroom can explain important aspects of the student's school experience. The teacher-student relationship, a central dimension of social capital, influences students' engagement, and the teaching style plays an important role in student outcomes. But there is scarce literature that links teaching styles to teacher-student relationship. This article aims to: 1) analyze whether there is a relationship between teaching styles and the type of relationship perceived by students; 2) test whether this relationship is equally strong for any teaching style; and 3) determine the extent to which students' perceptions vary according to their profile.

Design/methodology/approach:

A structural equation model with four latent variables is estimated: two for the teacher-student relationship (emotional vs. educational) and two for the teaching styles (directive vs. participative), with information for 21126 sixth-grade primary-students in 2019 in Spain.

Findings:

- Teacher-student relationships and teaching styles are interconnected.
- The participative style implies a better relationship.
- The perceptions of the teacher are heterogeneous, depending on gender (girls perceive clearer than boys) and with the educational background (children from lower educational background perceive both types of teaching styles more clearly).

Originality/value:

The analysis is based on the point of view of the addressee of the teacher's work, i.e. the student. It provides a model that can be replicated in any other education system.

The latent variables, based on a periodically administered questionnaire, could be estimated with data from diagnostic assessments in other countries, which in turn would allow the formulation of context-specific educational policy proposals that take into account student feedback.

## 1. Introduction

This article studies the student's perception of the teacher in the classroom. We consider the teacher as a "significant other" whose role should be analyzed in order to understand important aspects of the student's school experience, such as adequate integration, motivation and sense of belonging to the school, as well as their well-being, and resilience (Valdner, 2014; Van den Broeck, Demanet and Van Houtte, 2020; Anderson *et al.*, 2022). This role is analyzed on the basis of two dimensions. On one hand, the teacher-student relationship, which has been shown to be influential in students results (Van den Broeck, Demanet and Van Houtte, 2020). On the other, teaching styles, as these influence the classroom climate, which in turn plays an important role in student outcomes (Abello, Alonso-tapia and Panadero, 2020). There is also a still very scarce literature that underlines the importance of linking teaching styles and the teacher-student relationship, with particular emphasis on the relationship between less directive styles and positive social bonding of students at school (Opdenakker and Van Damme, 2006). The study of the relationship between students and teachers is approached in this paper from an unusual perspective, that of the student itself, determining the effect of teaching styles on the perceived relationship. In a context where the student's voice is of growing interest in the analysis of the learning process (Ralph, 2021), we believe it is important to propose models that provide insight into the student's perception, as the teacher's work.

This article therefore aims to: 1) analyze whether there is a relationship between teaching styles and the type of relationship perceived by students; 2) test whether this relationship is equally strong for any teaching style; and 3) determine the extent to which students' perceptions vary according to their profile. These three objectives are addressed by estimating the latent variables included in a structural equation model, which is estimated for the census database of the 2018/2019 Diagnostic Assessment of the Canary Islands (21126 students from 623 schools).

The structure of the article is as follows: Section 2 explains the theoretical background and describes the research hypotheses, Section 3 describes the features of the database and discusses the variable selection and the model, Section 4 describes the main results, Section 5 offers a discussion of the results, and Section 6 offers the main conclusions, including possible lines of research.

## 2. Theoretical framework

### 2.1. Classroom context as experiential context: teachers as significant others

The concept of "significant other" comes from symbolic interactionism and refers to those social actors who surround the subject and with whom, in their interaction, the subject shapes their own perception (Mead, 1972; Berger and Luckmann, 1995). In the analysis of educational processes, it implies underlining that the social character of education is determined both by macro-structures and by the specific contexts, such as the school communities, in which relational dynamics are produced (Brown, Daly and Liou, 2016). This perspective is in line with ecological and socio-cultural approaches that advocate the analysis of children's relational systems in order to understand their development (Mantzicopoulos and Neuharth-Pritchett, 2003). We thus speak of schools and classrooms as a relevant "experiential context" (Delamont, 1983; Elicker, 1997), where social capital is central for the student's experience (Daly, Liou and Der-Martirosian, 2021), and the role of the teacher in it. The teacher's performance in the classroom, as perceived by the student, is our point of reference to propose indicators that allow us to analyze both their relationship with the students and their teaching style, and the link between both dimensions.

**2.2. The interaction between teaching styles and teacher- student relationship.**

Although both teacher-student relationships and teaching styles have been analyzed separately, the interaction between the two dimensions is an underexplored area (Opdenakker and Van Damme, 2006). Research on this issue points that the two elements are interconnected. Opdenakker and Van Damme's analysis, focusing on the relationship between teaching styles and class management skills, concludes that both dimensions explain the presence of effective classroom practices (Opdenakker and Van Damme, 2006). Anderson et al. (2022) use a mixed method approach to analyze the interaction between students' involvement in their learning process (associated with more student-centered teaching styles), their meaningful relationships and their well-being. They conclude that "greater student participation is associated with greater wellbeing at school, while also pointing to the critical role of relationships of recognition in students' experiences of participation" (Anderson *et al.*, 2022).

In the field of Second Language Teaching, research has been conducted to test the hypothesis of a positive relationship between a participative teaching style and a better perceptions of teacher closeness and support. Findings from studies of Chinese native English learners suggest that the cultural context and the learning style play an important role in this positive relationship, as a more participative style may be perceived as stressful and may inhibit willingness to communicate in contexts where the teacher-directed model is more established (Rao, 2010). Also, when students are more self-conscious and insecure, a more participative style may imply a more negative perception of the teacher (Zhong, 2013).

Hence, evidence suggests that the interaction between teaching styles and teacher-student relationships may be very important for understanding student-wellbeing, as well as effective teaching (Chen, Dewaele and Zhang, 2022). Our study contributes to a better understanding of this interaction, which is still barely explored, by proposing indicators based on an existing instrument, with the aim of providing a model of analysis that would make it possible to follow up on this issue and make proposals for educational policy in this regard.

**2.3. The student's perspective**

While most studies analyze the role of the teacher and their interaction with students using independent observation (Mantzicopoulos and Neuharth-Pritchett, 2003; Slot *et al.*, 2017) or teacher questionnaires (Van Maele and Van Houtte, 2011; Van Houtte and Demanet, 2015), our model is based on the student's perspective, the addressee of the teacher's work.

The use of student perceptions as a form of feedback is considered to be a very interesting way to gain insight into the teacher's teaching quality (Rollett, Bijlsma and Röhl, 2021). However, our research does not focus on the quality of the teacher, but on the student's perceptions of the teachers teaching styles, and their perceptions of the teacher's educational and emotional relationship with the student. In any case, the Student Perceptions Questionnaires (SPQ) are a very promising way to collect student's feedback on their teachers' activities in the classroom and to provide teachers with useful feedback for their development (Röhl, Bijlsma and Rollett, 2021). Issues of validity and reliability cannot be ignored when interpreting the results (Bijlsma, 2021). The same can be said for the 'halo effects' of 'community' -or perceived teacher warmth- and/or student interest in the subject (Röhl and Rollett, 2021). Research also shows that perceptions of teachers change according to classroom characteristics (Fauth *et al.*, 2020) and the student's characteristics (Becker, 1952; Levy and Wubbels, 1992; Brandmiller, Dumont and Becker, 2020; Röhl, Bijlsma and Rollett, 2021).

Sortkær's research on students' perceptions of teacher feedback is relevant to this approach, as it describes the importance of teachers' actions actually being perceived as effective, and the need to acknowledge and identify students' characteristics in order to understand the different effects of teachers' practice (Sortkær, 2019).

### 2.4. Teacher-student relationship

Analyzing the relationship between teachers and students involves focusing on the links that are established between them in the context of classroom interaction. As Blumstein points out (2001) and Roseneil and Ketokivi (2016) will further elaborate, social bonds develop through the performance of roles in specific actions and in the transactions that result. In this respect, studies of teacher-student relations identify different dimensions. Based on Pianta's work (Pianta, 1994), some authors build up a three-dimensional model in which teacher closeness, the promotion of children's autonomy and conflict management are analyzed (Mantzicopoulos and Neuharth-Pritchett, 2003; Thijs and Fleischmann, 2015)

On the other hand, the work of Van Uden et al. highlights the importance of cognitive or educational attachment of teachers in fostering student engagement and a sense of belonging (Van Uden, Ritzen and Pieters, 2014). In a similar vein, the PISA studies, in relation to the role of teachers, analyze dimensions such as the educational relationship (named as teacher support), which is relevant in differentiating students' academic results. (OECD, 2019b). Other recent studies emphasize the positive influence of a close teacher-student relationship in academic achievement in adolescence (Magro *et al.*, 2023) and in moderating the effect of SES in mathematics achievement (Liu *et al.*, 2023).

### 2.5. Teaching styles

Teaching styles are the ways in which teaching activities that are carried out in the classroom can be classified with the aim of achieving particular learning outcomes. It is a concept that emphasizes, therefore, the instructional dimension of the teacher's role (Grasha, 1994; Socol, 2018). The literature indicates that different styles can lead to different student achievements (Opdenakker and Van Damme, 2006). Recent studies on fields such as sports or mathematics support this assertion (Mouratidou *et al.*, 2022; Villar-Aldonza, 2023). The classification by Mosston and Ashworth, who propose a "spectrum of teaching styles" (Mosston and Ashworth, 2008) based on the tasks carried out in the educational context, has been the basis for our proposal of indicators. Unlike models such as the Teaching Style Inventory, which is based on the values and strategies defined by the teacher (Grasha, 1994), the Mosston et al. model is based on the tasks carried out by the teacher. This feature allows to construct indicators based on the students' responses (Chatoupis, 2009). The spectrum categorizes the styles from the most directive (command style) to the one that provides greater autonomy and capacity for student participation (learner initiated) (Kulina and Cothran, 2003). It is a model widely used to analyze different aspects of the relationship between these styles and issues such as student satisfaction, enjoyment, and self-efficacy, especially in physical education (Chatoupis and Emmanuel, 2003; Fin *et al.*, 2019), but also in areas such as mathematics or language (Ngware, Mutisya and Oketch, 2012).

A similar classification is used by Reeve and Jang when contrasting teachers' "instructional behaviors", distinguishing between autonomy-supportive and controlling styles. It is important to emphasize that these distinctions do not analyze the perceived quality of the teacher, but rather the effect of, in Reeve and Jang's words, "what they say and do": their work in stimulating students' agency and intrinsic motivation (Reeve and Jang, 2006; Sheridan, Zhang and Konopasky, 2022), versus those that reflect the teacher's agenda as seen by the students.

### 2.6. Research hypothesis

The hypotheses formulated are based on a theory of action that emphasizes that teachers, when developing their activity in the classroom, display teaching styles that have a role in their relationships with their students. At the same time, students, as recipients of the teacher's activity, perceive these actions differently according to the actions displayed by the teachers, but also according to their characteristics. This means that what a student perceives as relevant is conditioned by his/her profile. How do different teaching styles influence the teacher-student relationship as perceived by students? How do the student's characteristics influence this perception? These are the main questions that foster our research. According to the literature reviewed, we can differentiate the directive teaching style (DirectiveTS) versus the participative teaching style (ParticipativeTS); just as we can differentiate the relationship between students and teachers as an exclusively pedagogical relationship (EducationalR) or as a more affective bond (EmotionalR). Our starting hypotheses are as follows:

H1: Teaching styles (directive and participative) have opposing effects on students' perception of the relationship.

> H1a: Directive style influences more the perception of the educational relationship than the perception of the emotional relationship (Fig. 1)

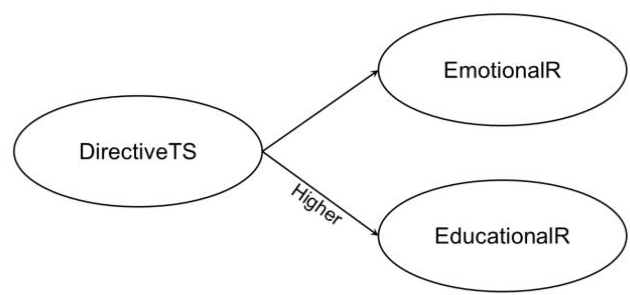

*Fig. 1. Hypothesis 1a*

H1b: Participative teaching style influences more the perception of the emotional relationship than the perception of the educational relationship (Fig. 2).

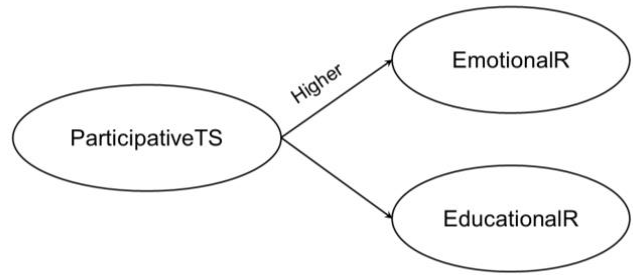

*Fig. 2. Hypothesis 1b*

H2: The directive teaching style generally implies a lower perceived relationship, both emotionally and educationally, than the participative style (Fig. 3).

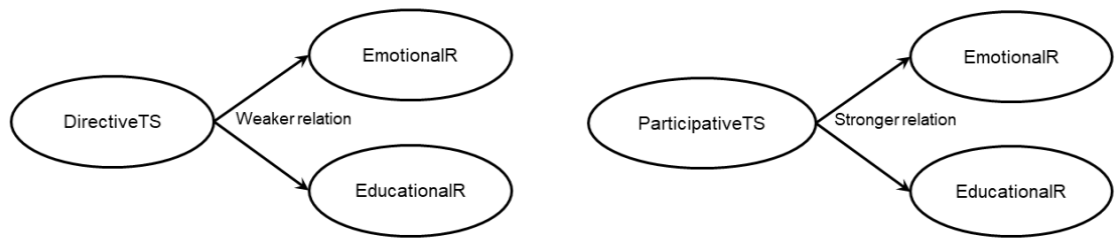

*Fig. 3. Hypothesis 2*

H3: The perception of the teaching styles and of the relationship vary according to student profile.

The research evidence for Hypotheses 1 and 2 is that an autonomy-supportive style is associated with perceptions of the teacher as more approachable and warm (Rao, 2010). However, there is also evidence that students' perceptions of their teacher are heterogeneous according to variables such as SES or gender (Becker, 1952; Levy and Wubbels, 1992; Brandmiller, Dumont and Becker, 2020), which is why we also propose Hypothesis 3.

## 3. Methodology

### 3.1. Database

The data used in this research has been provided by the Canarian Agency for University Quality Assurance and for Educational Assessment (ACCUEE), an autonomous body attached to the Ministry of Education of the Government of the Canary Islands. This institution is responsible for the annual implementation of the Diagnostic Assessment, with the aim of improving the Canarian education system. The evaluation consists of students taking tests to assess their competences (in linguistic communication, mathematics, science and technology and English) at different educational levels (3$^{rd}$ and 6$^{th}$ grade of primary education and 4$^{th}$ grade of secondary education) like in other international evaluations such as PISA, PIRLS or TIMSS. Context questionnaires are also administered to students, families, teachers, and school principals. This assessment can be census- or sample-based. The data we use in this research are those of the Diagnostic Assessment carried out in the 2018/2019 academic year, which is the last year available as a census. Specifically, it is the census of students enrolled in sixth grade of Primary Education in the Canary Islands' schools. In order to achieve the proposed objectives, we used exclusively the information from the 21126 records of the student questionnaire and, for some descriptive variables related to the socio-demographic profile, we used also the family questionnaire.

Table 1 presents the student profile according to gender, quarter of birth and educational level of both parents. The population of students in the 6$^{th}$ grade of primary education in the Canary Islands is slightly unbalanced by gender, with more boys than girls. The distribution by term of birth - as expected - is fairly equally distributed. Finally, the educational level of mothers is generally higher than that of fathers[1], although in both cases, most of them have upper secondary or lower tertiary education (ISCED 3-5) as their highest level of education.

*Table 1. Student Profile*

| Variable | Categories | Frequencies |
|---|---|---|
| Gender | Woman | 47.8 |
|  | Man | 52.2 |
| Quarter of birth | 1Q | 24.0 |
|  | 2Q | 23.1 |
|  | 3Q | 25.7 |
|  | 4Q | 27.2 |
| Mother's level of education | ISCED 0-1 | 9.7 |
|  | ISCED 2 | 21.8 |
|  | ISCED 3-5 | 41.6 |
|  | ISCED 6-7 | 26.2 |
|  | ISCED 8 | 0.8 |
| Father's level of education | ISCED 0-1 | 16.2 |
|  | ISCED 2 | 25.6 |
|  | ISCED 3-5 | 38.0 |
|  | ISCED 6-7 | 19.3 |
|  | ISCED 8 | 0.9 |

Note: Parents' levels of education are grouped according to the International Standard Classification of Education 2011 (Schneider, 2013): ISCED 0-1=Early childhood and Primary Education; ISCED 2=Compulsory Secondary Education; ISECD 3-5=Upper secondary and lower tertiary education; ISECD 6-7=Bachelor and Master degrees; ISECD 8=Doctoral degree.

Source: Own elaboration based on data from the Diagnostic Assessment 2018/2019.

### 3.2. Method

The context questionnaires do not have a single direct and specific question that asks about teaching styles, distinguishing between directive and participative styles. Nor is there a single question about the nature of the teacher-student relationship. Instead, there are many interrelated questions that address these unobservable variables of interest. In cases such as this, where our interest is focused on unobservable variables, estimating models using structural equation modeling (SEM) allows us to estimate these unobservable (latent) variables and even quantify the relationships (though not causality) between them (StataCorp, 2021).

---

[1] As indicated in the family questionnaire, the term "mother" refers to mother/legal guardian or first father/legal guardian in the case of male same-sex parent families. Similarly, the term "father" refers to father/legal guardian or second mother/legal guardian in the case of female same-parent families.

Thus, from the questions available in the student questionnaire, we selected the variables that characterize the teacher-student relationship and the teaching style. In order to identify the questions on the teacher-student relationship, we used the CLASS model (Slot *et al.*, 2017), and the CARTS (Vervoort, Doumen and Verschueren, 2015) and Y- CATS scales (Mantzicopoulos and Neuharth-Pritchett, 2003), as well as the Student Engagement Inventory questionnaire (Appleton *et al.*, 2006), and the PISA 2018 student questionnaire (OECD, 2019a). In terms of teaching styles, our reference model is Mosston & Ashworth's spectrum of teaching styles' (Mosston and Ashworth, 2008), as well as Reeve and Jang's distinction between autonomy-supportive and autonomy-thwarting teaching styles (Reeve and Jang, 2006). The specific selection of questions is set out in the Table 2.

The definition and operationalization of the latent variables began with a list of the most relevant variables in the cited questionnaires, and, in addition, the identification of the relationship between the variables and the theoretical framework. Since our questionnaire was not designed ad hoc, but for a more general purpose, questions similar to those in the cited instruments were identified and grouped according to two categories: teaching styles and teacher-student relationship, theoretically defined by distinguishing between the activities used by the teacher in the classroom to teach, on the one hand, and the relationship perceived by the student, on the other. In addition, through a confirmatory factor analysis, a clearly definable distinction was found between the directive and participative teaching styles on the one hand, and between the emotional and educational bond on the other.

The latent variable Directive Teaching Style (DirectiveTS) reflects an instructional strategy in which the teacher is the protagonist of the process, indicating the tasks to be carried out to complete the learning process, and a predominance of frontal teaching (teacher explaining, student taking notes). The teacher's agenda is carried out. The latent variable Participative Teaching Style (ParticipativeTS) is characterized by an active role of the student, who participates in decisions that affect the group, contributes to the learning process, and works cooperatively, thus reporting not only that teachers facilitate an active role of the student, but also a sensitivity to student's needs and efforts (Reeve and Jang, 2006). On the other hand, the latent variables referring to the relationship between the student and the teacher refer to emotional and educational bonding. EmotionalR relates to the perception of respect, fair treatment and a positive classroom climate. While the educational relationship (EducationalR) is related to the clarity of the teacher's presentation, appropriate feedback and interest in the proposed tasks (Mantzicopoulos and Neuharth-Pritchett, 2003).

*Table 2. Questions that make up each latent variable*

| | | |
|---|---|---|
| | DIRECTIVE TEACHING STYLE (DirectiveTS) | |
| | v2019a5b | We present works or topics |
| | v2019a5c | As they explain, we are asked about the issues |
| TEACHING | v2019a5e | We hold debates in class |
| STYLE | v2019a5i | We take notes |
| | v2019a5l | We study individually |
| | PARTICIPATIVE TEACHING STYLE (ParticipativeTS) | |
| | v2019a9b | Students participate in decisions (rules, outings, etc.) |
|   | v2019a10g | My classmates help me in class |
|   | v2019a12g | My teachers let me demonstrate what I have learnt |
|   | v2019a13g | My teachers take into account the grade we give each other |
|   | v2019a13h | My teachers value interest and participation in class |
| RELATIONSHIP | **EMOTIONAL RELATIONSHIP (EmotionalR)** | |
|   | v2019a9a | Teacher is happy with the group |
|   | v2019a9c | I am respected and feel safe in my class |
|   | v2019a9d | I like the way my classroom is organized and decorated |
|   | v2019a11e | Most of my teachers treat me fairly |
|   | v2019a12j | My teachers listen to what I have to say |
|   | **LEARNING RELATIONSHIP (EducationalR)** | |
|   | v2019a9e | I really like the work I do in the classes |
|   | v2019a12a | I know what my teachers expect me to do |
|   | v2019a12b | It is easy to understand my teachers |
|   | v2019a12c | I'm interested in what my teachers say |
|   | v2019a12d | My teachers suggest interesting things for me to do |
|   | v2019a12e | My teachers answer my questions with clarity |
|   | v2019a12f | My teachers are good at explaining |
|   | v2019a12i | My teachers tell me how to improve when I make mistakes |

Note: All questions are Likert-type questions. The questions related to the frequency of an action or situation ranged from the absence to the systematic presence of the action or situation (Never - Almost never - Almost always - Always), while the response categories related to the degree of agreement were as follows: Not at all agree - Somewhat agree - Fairly agree - Strongly agree.

Source: Own elaboration based on the student questionnaire of the Diagnostic Assessment 2018/2019.

Once the questions have been selected, they are interrelated using structural equation modelling (Fig. 4), in which the latent variables of interest are estimated by maximum likelihood (e.g. ovals in the Fig. 4), as well as the interrelationships of teaching styles with perceptions of the emotional and educational relationship (e.g. black arrows of the Fig. 4). In addition, in order to be able to estimate the specific value of the latent variables for each student, the value of each of them is estimated by means of Confirmatory Factor Analysis.

The estimation of the structural equation model allows to test hypotheses 1 and 2. For this purpose, it is only necessary to compare the value of the standardized coefficients (values A, B, C and D in Fig. 4) of the structural equation model estimated by maximum likelihood.

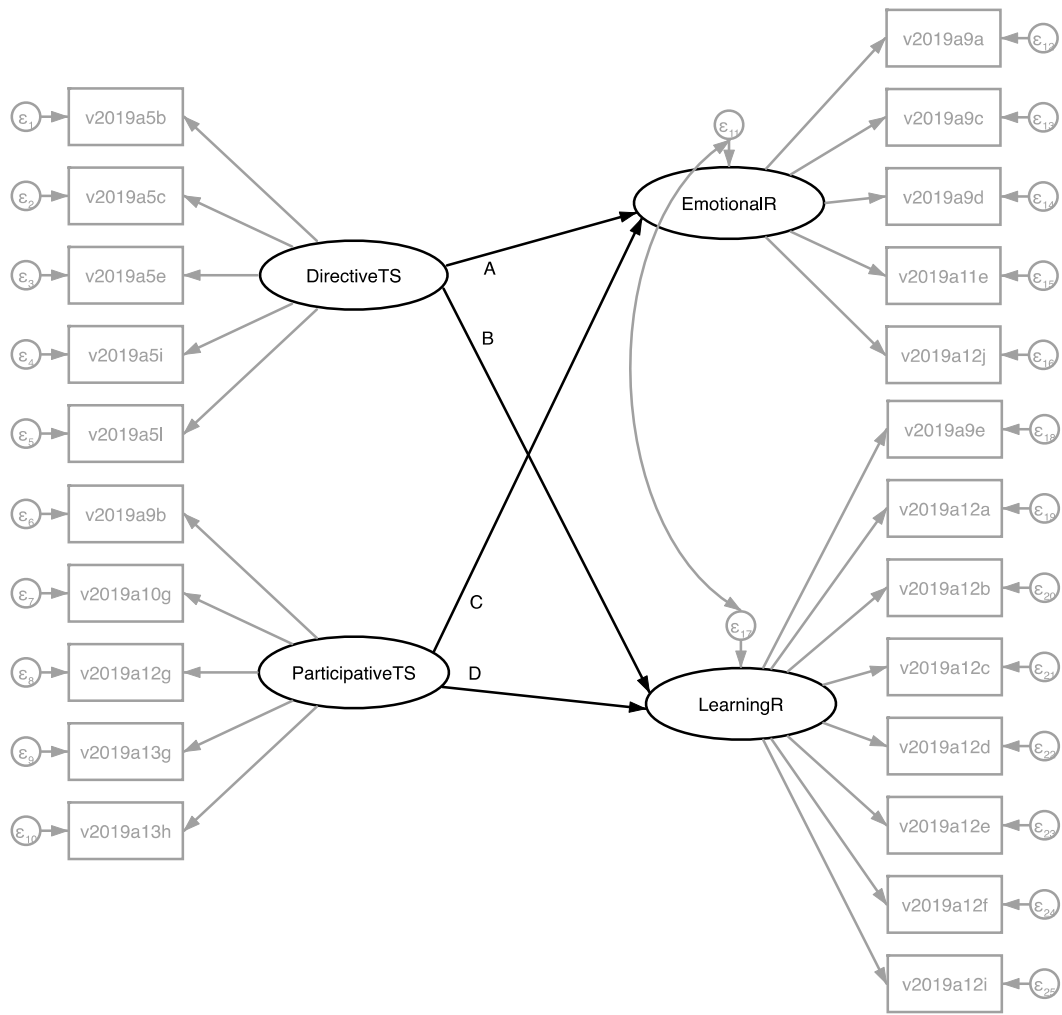

*Fig. 4. Relationship between teaching styles and perceptions of teacher-student relationship (theoretical SEM)*

Note: Hypothesis H1a is satisfied if B > A; H1b is satisfied if C > D; H2 is satisfied if (C and D) > (A and B).

Finally, hypothesis 3 is tested by carrying out various mean-comparison tests (t-test or oneway analysis-of-variance), according to the characteristics of the Table 1, for each of the latent variables (DirectiveTS, ParticipativeTS, EmotionalR, EducationalR) obtained through confirmatory factor analyses.

All statistical and econometric analysis was performed with StataSE 17 (StataCorp, 2021).

### 4. Results

#### 4.1. SEM

The SEM model that allows us to test hypotheses 1 and 2, presents a good fit [$\chi^2_{(225)}$=9847.865, p=0.000; RMSE< 0.05, p=0.811; CFI = 0.897; TLI= 0.884; $R^2_{EmotionalR}$=0.947; $R^2_{LearningR}$=0.871; $R^2_{Overall}$= 0.978]. All estimated coefficients are significant at 1% and all have a positive sign (Fig. 5), thus inferring a positive correlation between teaching styles and perceived relationship. Furthermore, given the values of the estimated coefficients, Hypothesis 1 and Hypothesis 2 are confirmed. Therefore, we can state that: 1) the directive teaching style has more influence on the perception of the educational relationship, while the participative teaching style has more influence on the perception of the emotional relationship; and 2) in any case, the participative teaching style correlates much more with the perception of any type of relationship than the directive teaching style.

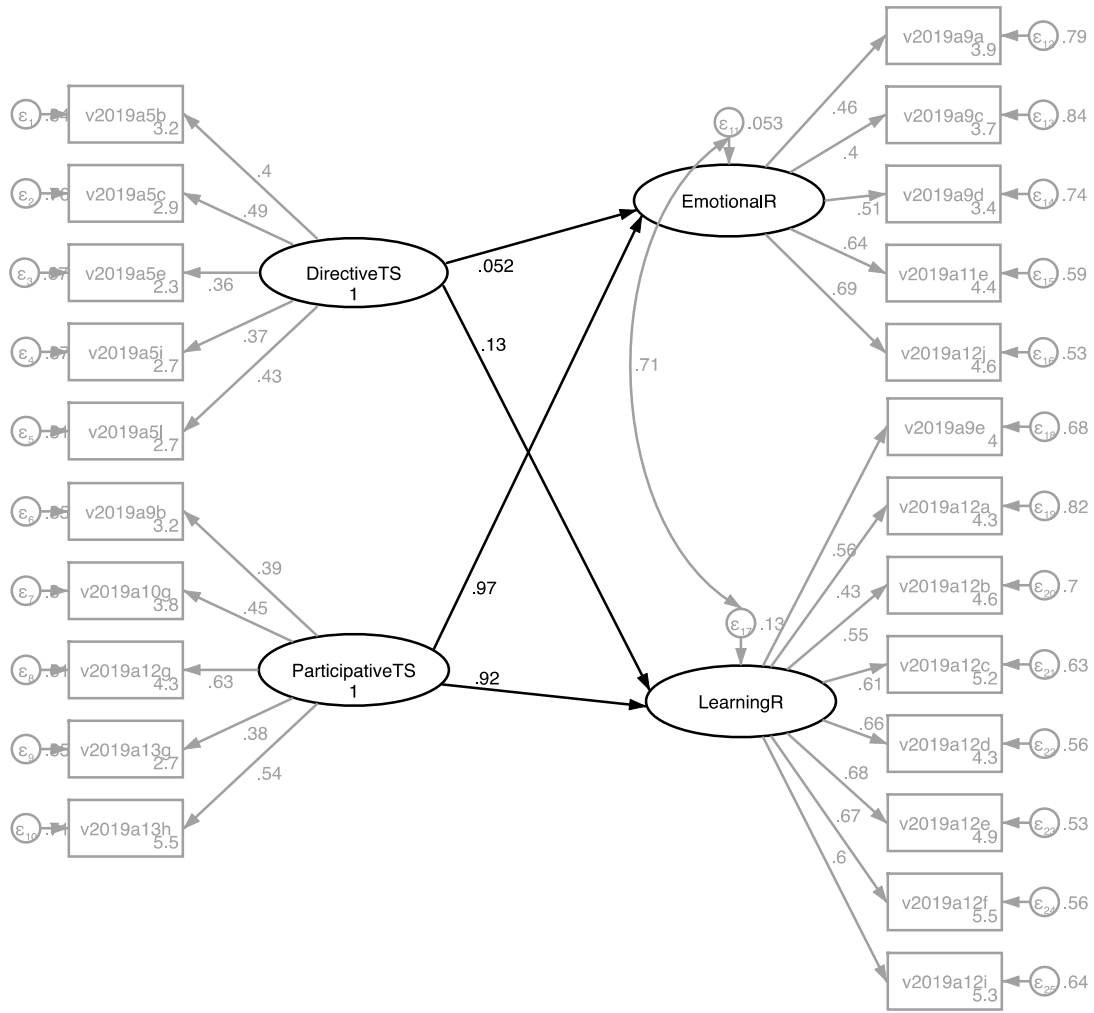

*Fig. 5. Relationship between teaching styles and perceptions of teacher-student relationship (Estimated SEM)*

### 4.2. Mean-comparison tests

Once the four latent variables have been estimated by confirmatory factor analysis, we proceed to perform mean-comparison tests to see if there are significant differences in the perception of students according to gender (Table 3), their relative age (Table 4) and the level of education of their parents (Table 5 and Table 6), which allows us to test Hypothesis 3.

Boys and girls show no differences in their perception of the directive teaching style (Table 3). However, there are differences between them in the perception of the participative teaching style (t=6.004, p=0.000), and in the type of relationship (Emotional: t=6.682, p=0.000; Educational: = 5.352, p=0.000). In fact, girls take higher values than boys in these last three latent variables, so they perceive a greater use of the participative style than boys, and they perceive a greater use of the emotional and learning relationship than boys.

*Table 3. Perception according to gender*

| Latent variables | Girls | Boys | t (p) | d Cohen | 95% CI |
|---|---|---|---|---|---|
| DirectiveTS | -0.01 (0.70) | 0.01 (0.71) | -1.623 (0.105) | -0.025 | -0.054  0.005 |
| **ParticipativeTS** | 0.04 (0.76) | -0.03 (0.78) | 6.004 **(0.000)** | 0.091 | 0.061  0.121 |
| **EmotionalR** | 0.04 (0.82) | -0.04 (0.86) | 6.682 **(0.000)** | 0.101 | 0.071  0.130 |
| **EducationalR** | 0.04 (0.90) | -0.04 (0.93) | 5.352 **(0.000)** | 0.081 | 0.051  0.110 |

Note: Variables in bold have differences in means

Regarding the quarter of birth (Table 4), no differences are observed for the latent variables analyzed, so that the relative maturity of 6th graders does not seem to be a determining factor in the perception of teaching style or the teacher-student relationship.

*Table 4. Perception according to quarter of birth*

| Latent variables | 1st quarter (a) | 2nd quarter (b) | 3rd quarter (c) | 4th quarter (d) | F (p) | Tukey (p<0,1) |
|---|---|---|---|---|---|---|
| DirectiveTS | -0.01 (0.70) | -0.02 (0.68) | -0.01 (0.71) | -0.02 (0.71) | 0.33 (0.81) | - |
| ParticipativeTS | 0.01 (0.78) | 0.01 (0.76) | 0.02 (0.75) | 0.00 (0.76) | 0.14 (0.936) | - |
| EmotionalR | 0.01 (0.83) | 0.01 (0.82) | 0.01 (0.84) | 0.02 (0.84) | 0.06 (0.980) | - |
| EducationalR | 0.01 (0.91) | 0.00 (0.91) | 0.02 (0.89) | 0.02 (0.90) | 0.74 (0.526) | - |

Note: Variables in bold have differences in means

Differences are observed in all latent variables when differentiating students by the educational level of their mothers (Table 5), except for the emotional relationship variable (F=0.93, p=0.443). In all cases where there are statistically significant differences, the corresponding latent variable takes a higher value for those with low educated mothers (e.g. ISCED0 to ISCED2), compared to those with higher educated mothers (e.g. ISCED3 to ISCED7), as shown in the column with the Tukey contrast of the Table 5. The group of mothers with ISCED8 education level is quite small, and the differences with it are not statistically significant.

*Table 5. Perception according to Mother's Education Level*

| Latent variables | ISCED0-1 (a) | ISCED2 (b) | ISCED3-5 (c) | ISCED6-7 (d) | ISCED8 (e) | F (p) | Tukey (p<0,1) |
|---|---|---|---|---|---|---|---|
| **DirectiveTS** | 0.10 (0.70) | 0.07 (0.71) | -0.01 (0.70) | -0.06 (0.71) | -0.03 (0.65) | 17.34 **(0.000)** | a/c; a/d; b/c; b/d; b/d; c/d |
| **ParticipativeTS** | 0.08 (0.75) | 0.07 (0.76) | 0.03 (0.73) | 0.01 (0.77) | 0.15 (0.73) | 3.58 **(0.006)** | a/d; b/d |
| EmotionalR | 0.08 (0.82) | 0.06 (0.82) | 0.05 (0.81) | 0.04 (0.84) | 0.15 (0.79) | 0.93 (0.443) | - |
| **EducationalR** | 0.10 (0.91) | 0.10 (0.88) | 0.04 (0.88) | 0.02 (0.89) | 0.10 (0.87) | 3.88 **(0.000)** | a/d; b/c; b/d |

Note: Variables in bold have differences in means

In the case of differentiating by parents' educational level (Table 6), the only latent variable that shows statistically significant differences is the perception of directive teaching style (F=8.58, p=0.000). As with the mothers' educational level, those who have fathers with a lower educational level (e.g. ISCED0 to ISCED2), perceive more directive teaching style than those who have fathers with a higher educational level (e.g. ISCED6 to ISCED7), as shown in the column with the Tukey's test for the Table 6. The rest of the latent variables show no significant differences by parents' level of education.

*Table 6. Perception according to Father's Education Level*

| Latent variables | ISCED0-1 (a) | ISCED2 (b) | ISCED3-5 (c) | ISCED6-7 (d) | ISCED8 (e) | F (p) | Tukey (p<0.1) |
|---|---|---|---|---|---|---|---|
| **DirectiveTS** | 0.08 (0.71) | 0.02 (0.72) | 0.00 (0.69) | -0.05 (0.69) | -0.03 (0.76) | 8.58 **(0.000)** | a/c; a/d; b/d; c/d |
| ParticipativeTS | 0.07 (0.75) | 0.06 (0.74) | 0.03 (0.74) | 0.01 (0.77) | 0.04 (0.76) | 1.97 (0.096) | - |
| EmotionalR | 0.07 (0.83) | 0.05 (0.82) | 0.05 (0.81) | 0.07 (0.81) | 0.07 (0.86) | 0.38 (0.826) | - |
| EducationalR | 0.09 (0.91) | 0.08 (0.87) | 0.05 (0.88) | 0.02 (0.89) | 0.06 (0.81) | 1.98 (0.095) | - |

Note: Variables in bold have differences in means

Therefore, Hypothesis 3 is also confirmed, since both the teaching style and the relationship are perceived differently according to the characteristics of the student, at least according to the student's gender and the educational level of his or her parents, if not according to their relative age.

## 5. Discussion

The results presented above confirm the relevance of analyzing the relationship between teaching styles and the teacher-student relationship in the analysis of the classroom context (Hypotheses 1 and 2). Thus, the study by Goldman & Goodboy (2014) points out the importance of analyzing both teachers' interactions with students and their teaching styles in order to understand the emotional experiences of college students in the classroom. Chen et al. (2022) stress the importance of both teaching style and the affective relationship between teachers and students for students' willingness to communicate in introductory second language classes.

Likewise, research supports the enhancing effect of more participative styles on the positive bond between students and teachers. This is the case of the study by Wang et al. (2016) on secondary school students in Singapore, who found that an autonomous supportive teaching style improved students' perception of the satisfaction of their basic needs in the classroom. Zee & Koomen (2020) identified, in 23 regular Elementary schools in the Netherlands, that there was a positive association between teaching strategies in relation to student autonomy and students' perception of closeness in their affective relationship with teachers.

Regarding the student profile, we observe that certain student characteristics, such as student gender and parental education level, are sensitive to both the relationship with the teachers and teaching styles, but others, such as student relative age, are not (Hypothesis 3).

*In terms of gender,* girls perceive the participative style more, as well as the emotional and educational relationships. It is noteworthy that, to the best of our knowledge, there are no studies that relate teaching styles to gender, but they do relate to teacher-student interaction. In this regard, our results are consistent with existing research on gender differences and social interactions, which emphasize that girls perceive more clearly their relationship, both affective and educational, with their teachers, and establish closer relationships with them (Rueger, Malecki and Demaray, 2008; Rautanen *et al.*, 2021). Hamre and Pianta's studies indicate that this assessment is bidirectional, and that teachers also perceive a closer relationship with female students (Hamre and Pianta, 2001). Several studies have proven the relationship between positive teacher-student bonds and the student's adjustment and involvement in school (Rueger, Malecki and Demaray, 2008; Tennant *et al.*, 2015; Havik and Westergård, 2020; Rautanen *et al.*, 2021), and even school outcomes (Hamre and Pianta, 2001; Agasisti *et al.*, 2021). This has led some authors to suggest that these differences in the relationship may partly explain the poorer performance of boys during their compulsory schooling (Van Houtte, 2020).

*As for the quarter of birth*, our results show that there are no differences in the perception of teaching styles, nor of the teacher-student relation, motivated by a difference in maturity between students, not even between students born in the first quarter and those born in the last quarter of the year. Therefore, although the literature has noted the influence of this maturational gap on the educational performance of primary school students (Verachtert *et al.*, 2010; González-Betancor and López-Puig, 2015b, 2015a; Bjerke *et al.*, 2021), this gap does not seem to influence their perception of the relationship they have with their teachers.

*With regard to the parents' educational level*, while differences can be seen with regard to mothers' educational level in almost all the dimensions analyzed, fathers' educational level only shows significant differences in terms of directive style. In this sense, the results are consistent with the greater weight of mothers' education in other aspects of the school experience, such as educational achievement (Rodríguez-Rodríguez and Guzmán, 2021).

*In relation to teaching styles*, we identified that the lower the mother's and father's level of education, the higher the perception of the directive style. There are differences between practically all categories. It is, therefore, a style very clearly perceived by the students, and more clearly perceived the lower the educational background. This may indicate that the teacher's strategies and behaviors change according to the type of cultural capital they believe their students have, employing a more directive style with those students they perceive as unequal. The ethnographic study by Palludan (2007) analyzes this question, drawing on Bourdieu's theory of habitus and Bernstein's theory of linguistic codes. By comparing the language used in the classroom by kindergarten teachers when addressing students of Danish and non-Danish origin, the author highlights the performance of two "teaching tones" by teachers: a) an instructional teaching tone, with which teachers address children of non-Danish origin, and b) a more symmetrical and conversational teaching tone, used for children of Danish origin (Palludan, 2007). But it is also possible that such a directive style, which implies a greater teacher presence in the teaching process, is perceived more clearly by students with lower cultural capital. In this respect, Sortkær's work on the student's perception of teacher feedback according to SES in the Nordic countries, which also draws on Bourdieu and Bernstein, points out that the more controlling style, characterized by explicit guidance of the student in learning, is perceived more clearly by students with low SES in Denmark, but not in the rest of the Nordic countries (Sortkær, 2019).

With regard to the participative style, significant differences in perception can be seen between students whose mothers have at most secondary education and those who have master's and bachelor's degrees, but not between them and those whose mothers have doctoral degrees. Specifically, there is a greater perception of this style among students whose mothers have the lowest level of education. In this respect, our study presents opposite results to Sortaeker's, which suggests that students with higher SES clearly identify facilitative feedback, characterized by providing more autonomy to students in their learning (Sortkær, 2019). Our results, in this sense, provide a basis for the need to analyze more specifically the general hypothesis of high cultural capital and the automatic internalization of a habitus centered on self-regulation (Edgerton, Roberts and Peter, 2013).

As far as the teacher-student relationship is concerned, differences in the perception of the learning relationship are only observed regarding the mother's level of education. It is interesting to note that the contrast in perceptions occurs between mothers' lower educational levels and bachelor's and master's degree level. Moreover, the most positive perception of the learning relationship occurs for students whose mothers have the lowest educational categories. On the other hand, again, there are no significant differences in this perception in relation to the sons and daughters of mothers with doctoral studies.

Previous studies (Becker, 1952; Brandmiller, Dumont and Becker, 2020) highlight the poorer relationship between students of lower social status and their teachers, due to differences in social background and cultural capital. However, these results are based on teachers' opinions and behaviors without differentiating the typology of the teacher-student relationship. Our study complements this literature by contributing for the first time, to the best of our knowledge, the student's perspective and, above all, by differentiating the type of emotional relationship from the educational one. For this reason, our results suggest that the educational relationship between students and their teachers is more highly valued among students whose mothers have lower levels of education than among students whose mothers have higher levels of education (see in Table 5 the difference in means for students whose mothers have ISCED 0 to 2, compared to the means for those whose mothers have ISCED3 to 7). However, the emotional relationship seems to be independent of social background, since there are no statistically significant differences in the means by relative age or parental education level (Table 4 to Table 6), although there are significant differences by gender (Table 3).

Our results also indicate that, although fostering student autonomy and participation in learning decisions (ParticipativeTS) implies a higher valuation of the affective educational relationship (because of the higher coefficients in Fig. 5, which confirm Hypothesis 2), the cultural capital of origin plays an important role in the perception of this relationship, since, in this case, a greater clarity in discerning the directive style (DirectiveTS) does not imply a worse perception of the educational relationship either (since the estimated coefficients in Fig. 5 are also positive, although lower than those of ParticipativeTS). In this respect, and in the absence of studies to corroborate these results, we can suggest that it is quite possible that, for students with a low family educational background, a directive teaching style is valued as a positive educational relationship, insofar as it implies a clear presence of the teacher in the classroom and can be interpreted as a sign of interest and involvement.

## 6. Conclusions

The main objective of this research was to provide a model to analyze the role of the teacher in the classroom as a "significant other", identifying and relating their teaching styles and the teacher-student relationship, both educational and affective. Although these elements have traditionally been analyzed separately, we have proven the relevance of doing so in a combined way, thus being able to identify the higher correlation of the participative style versus the directive style in the emotional and educational relationship.

Likewise, the mean-comparison tests have allowed us to verify that the students' perception is heterogeneous, and that both gender and the educational background of the family (especially the mother's) configure differentiated perceptions with respect to teaching styles and the relationship between teachers and students. In terms of gender, the results on the perception of affective and educational interactions more clearly perceived by girls are in line with the existing literature. With regard to the perception of teaching styles, despite not having identified previous studies in this respect, our results indicate that there are no significant gender differences in the perception of the directive style, but there are significant differences in the participative style, which girls perceive more clearly.

The analysis of differences by educational background is novel and shows a more complex picture. The results obtained complement the literature on family background and students' perceptions of teachers. It is particularly striking to note the lower identification of a participative style by students whose mothers have a higher level of education. It is also interesting to note that the lower the mother's level of education, the greater the identification of the educational bond. These findings suggest that the perception of teacher behavior in the classroom is very different according to family cultural capital, and it is a line of research that may be very fruitful in understanding the role of different types of teachers in students' aspirations, engagement, and achievement, as well as the effect of teaching strategies in students with very unequal cultural capital. The results ultimately show that the heterogeneity of society in terms of cultural capital and gender is replicated in the classroom in the form of students' equally heterogeneous perceptions of the teacher's role. As for the teacher, as potential recipient of the student's feedback, our results emphasize that a better understanding of teaching practice, as developed in the models of student feedback on teaching (Röhl, Bijlsma and Rollett, 2021), requires the analysis of the classrooms' internal diversity, which means different needs and expectations. Teaching is a contextualized practice that can greatly benefit from knowledge of the particularities of students' perceptions in order to facilitate dialogue about teaching and a more "informed practice" by the teachers (Jones and Hall, 2021).

The main limitation of the study has to do with working with a pre-existing questionnaire, which was not designed for the purpose of the present study. For this reason, in the configuration of the teacher-student relationship, the behavioral relationship, which concerns discipline and conflict management by teachers (Mantzicopoulos and Neuharth-Pritchett, 2003; Vervoort, Doumen and Verschueren, 2015; Slot *et al.*, 2017), could not be analyzed, as the questionnaire lacked the appropriate variables. Similarly, the teaching styles had to be constructed in a model of oppositions instead of a spectrum (Chatoupis, 2009), as we would have preferred. Nevertheless, the proposed model has many advantages: (a) the chosen estimation method (SEM) allows the estimation of the latent variables of interest; (b) this methodology also allows measuring the teacher-student relationship through these latent variables from the student's point of view; (c) it is contextualized at the primary education stage (like TIMSS and PIRLS), which allows detecting relationships at the beginning of the educational process; d) the information comes from a questionnaire applied to a census of students and not from direct observation, which increases the amount of information; e) it can be reproduced periodically, since the evaluation of the Canarian education system is carried out periodically by ACCUEE, which makes it possible to see the evolution of the patterns detected; and f) it can be reproduced in any other context (country, educational level), as long as similar questions are used.